\documentclass[apalike]{article}
\usepackage{graphicx}
\usepackage{amsmath}
\usepackage{natbib}
\begin{document}

\begin{center}
\Large \textbf{Quantum physics meets biology}\\
\vspace{0.5cm}
\large Markus Arndt$^1$, Thomas Juffmann$^1$ and Vlatko Vedral$^{2,3}$\\
\vspace{0.2cm}
\small $^1$ Faculty of Physics, University of Vienna, Boltzmanngasse 5, 1090 Vienna, Austria \\
$^2$  Atomic and Laser Physics, Clarendon Laboratory, University of Oxford, Parks Road, Oxford OX1 3PU, UK\\
$^3$ Department of Physics + Centre for Quantum Technologies, National University of Singapore, 2 Science Drive 3, Singapore 117543
\end{center}

\begin{abstract}
Quantum physics and biology have long been regarded as unrelated
disciplines, describing nature at the inanimate microlevel on the one hand
and living species on the other hand. Over the last decades
the life sciences have succeeded in providing ever more and
refined explanations of macroscopic phenomena that were based on
an improved understanding of molecular structures and mechanisms.
Simultaneously, quantum physics, originally rooted in a world
view of quantum coherences, entanglement and other non-classical
effects, has been heading towards systems of increasing
complexity. The present perspective article shall serve as a
pedestrian guide to the growing interconnections between the
two fields. We recapitulate the generic and sometimes unintuitive
characteristics of quantum physics and point to a number of
applications in the life
sciences. We discuss our criteria for a future
quantum biology, its current status, recent experimental progress and also the restrictions that nature imposes on bold extrapolations of quantum theory to macroscopic phenomena.

\end{abstract}

\section{Introduction}
While in the days of Darwin and Mendel the life sciences were
mainly focusing on botany or zoology, modern biology,
pharmacology and medicine are deeply rooted in a growing
understanding of molecular interactions and organic information
processing.

Quantum physics, on the other hand, was initially centered on microscopic phenomena with photons,
electrons and atoms. But objects of increasing complexity have
attracted a growing scientific interest, and since the size
scales of both physics and the life sciences have approached each
other, it is now very natural to ask: What is the role of quantum
physics in and for biology?

Erwin Schr\"{o}dinger, most famous for his wave equation for
non-relativistic quantum mechanics, already ventured across the
disciplines in his lecture series {\em What is life?}~(\citealp*{Schroedinger1944a}). He anticipated a molecular
basis for human heredity which was later confirmed to be the DNA molecule ~(\citealp*{Watson1953}).
Since the early days of quantum physics its influence on biology has always been present in a reductionist
sense: quantum physics and electrodynamics shape all molecules and thus
determine molecular recognition, the workings of proteins and DNA.
Also van der Waals forces, discrete molecular orbitals, the stability of matter: all this is quantum physics and a natural basis for life and
everything we see.

But even a hundred years after its development, quantum physics is
still a conceptually challenging model of nature: it is often
acclaimed to be the most precisely verified theory of nature and
yet its common interpretation stands in discrepancy to our
classical, i.e. pre-quantum, world view and our natural ideas about reality or
space-time. Is there
a transition between quantum physics and our
every-day world? And how will the life sciences then
fit into the picture - with objects covering anything from molecules
up to elephants, mammoth trees or the human brain?

Still half a century ago, the topic had some
rather skeptical reviews~(\citealp*{LonguetHiggins1962}). But experimental advances have raised a new awareness and several recent
reviews (e.g.~\citealp*{Abbot2008}) sketch a more optimistic picture that may be over-optimistic in some aspects.

The number of proven facts is still rather
small. Many hypotheses that are formulated today may be found to
have lacked either visionary power or truth, by tomorrow.
We will therefore start on well-established physical grounds and
recapitulate some typical quantum phenomena. We will then elucidate
the issue of decoherence and dephasing, which are believed to be central
in the transition between the quantum and the classical world. They are often regarded to be the limiting factors if we want to observe quantum effects on the macroscopic scale of life. Next, we give an overview over modern  theories and experiments at the interface between quantum physics and biology.
The final section will be devoted to open speculations, some of which are still facing less supporting experimental evidence than theoretical counterarguments.

Experimental studies at the interface between quantum physics and the life sciences have so far been focused on two different questions:
\begin{itemize}
\item
{\em Can genuine quantum phenomena be realized with
biomolecules?} \\Photon
anti-bunching in proteins~(\citealp*{Sanchez-Mosteiro2004}), the
quantum delocalization of biodyes in matter-wave
interferometry~(\citealp*{Hackermuller2003a}) and the
implementation of elementary quantum algorithms in
nucleotides~(\citealp*{Jones1999}) are some
recent examples.\\
These experiments are optimized for revealing fundamental
physics, such as quantum statistics, delocalization and
entanglement.
But they all also show that quantum phenomena are best observed in
near-perfect isolation from the environment or at ultra-low
temperatures, in order to avoid the detrimental influence of decoherence and dephasing.  They are thus not representative for life as such.
\item
{\em Are non-trivial quantum phenomena relevant for life?}\\
Non-trivial quantum phenomena are here defined by the presence of
long-ranged, long-lived or multi-particle quantum coherences, the
explicit use of quantum entanglement, the relevance of single
photons or single spins triggering macroscopic phenomena.\\
Photosynthesis, the process of vision, the sense
of smell or the magnetic orientation of migrant birds are
currently hot topics in this context. In many of these cases the
discussion still circles around the best interpretation of
recent experimental and theoretical findings.
\end{itemize}

\section{A brief review of elementary quantum phenomena}
Quantum physics includes a wide variety of phenomena.
Most of them are regarded as unusual because they violate our everyday
expectations of how nature should behave.

\subsection{Quantum discreteness}
Quantum physics derives its very name from the discreteness of
nature. The latin word {\em quantum} asks the question {\em how much?} and
even colloquially a quantum is nowadays a {\em small portion}.
Several physical properties only assume a countable number of
values. This is for instance true for the electronic energy
in atoms or the vibrational energy in molecules.
The quantized set of spectral lines, Fig.~\ref{Quantum}a), is often used in the life sciences as a finger print of chemical substances, since it relates to the discrete set of energies in all nanomatter, from atoms to large biomolecules.  And yet, this aspect is regarded as atomic and chemical physics rather than a part of quantum biology.
If, however, single photons or single spins can trigger a chain
of macroscopic phenomena in organic systems the assignment will be justified.

\subsection{Quantum superposition}
One of the intriguing aspects of physics is related to the
fact that quantum states are well-defined even when they refer to
situations which we would describe as a coexistence of mutually
excluding possibilities in classical physics.
We usually adhere to Aristotle's sentence of non-contradiction, that
the same attribute cannot at the same time belong and not belong to the same
subject. Quantum physics teaches us that we
either have to give up this wisdom or otherwise
renounce on other well-established intellectual concepts, such as
our understanding of reality, information or space-time.

\subsubsection{Wave-particle duality of light}
\begin{figure} [htb]
 \begin{center}
 \includegraphics[width=0.8\columnwidth]{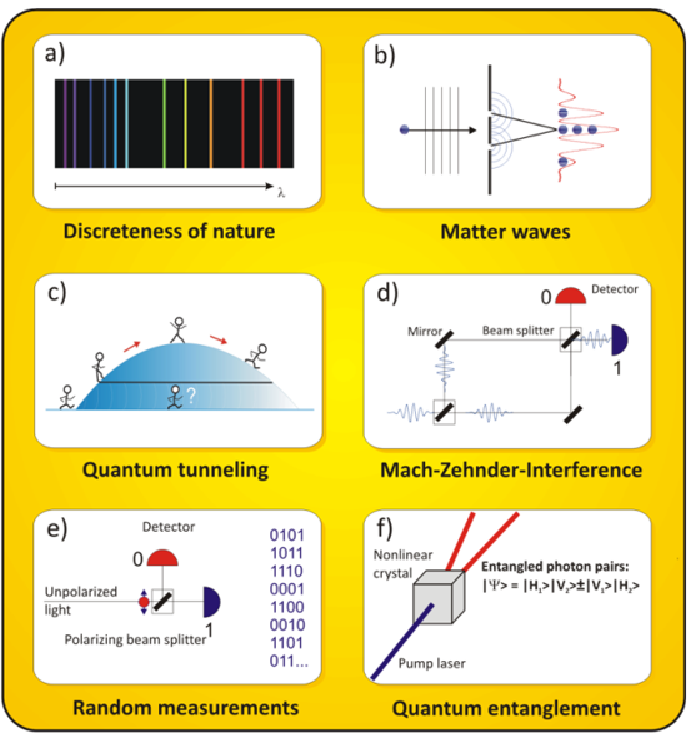}
  \caption{Review of some prominent quantum phenomena: a) Quantum physics emphasizes that our world is built on discrete particles that are bound in finite systems of discontinuous energies. This becomes evident in the finite number of wavelengths $\lambda$, respectively colors, that an atom emits.
(b) The quantum delocalization and  wave-particle duality of light and matter can be demonstrated using a double-slit experiment (see text).   (c) Quantum wave effects allow tunneling through an energy barrier which would classically be insurmountable. (d) A Mach-Zehnder interferometer allows to split a wave into two widely separated paths. (e) A quantum measurement generates objective randomness. A photon beam, that is divided by a 50/50 beam splitter, will hit either detector randomly, and yield an absolutely random sequence of zeros and ones. (f) Entanglement is the inseparable quantum correlation of two or more  particles or degrees of freedom. Here, we sketch the creation of a polarization entangled pair of red photons when a single uv pump photon interacts with a nonlinear crystal~(\citealp*{Kwiat1995a}). A measurement of both photons shows perfectly anti-correlated polarizations although the result on each side individually appears to be absolutely random.
  \label{Quantum}}
 \end{center}
 \end{figure}

A paradigmatic example for this fact is the dual nature of light, that manifests itself in Young's famous double-slit diffraction
experiment (Fig.~\ref{Quantum}b): When a light beam is
sent onto a small opening in a wall, it will spread out
behind the slit. If we open a neighboring slit in the screen, we will observe a fringe pattern (Fig.~\ref{Quantum}b) that exhibits dark minima at
several screen positions that are still bright when only a single
slit is open. This observation is explained by the
superposition and interference of classical electromagnetic
waves of well-defined relative phases i.e. sufficient mutual
coherence.

The same wave interference experiment turns into a quantum puzzle when we dilute the optical field to the level of individual quanta of light, photons, which are always detected as localized packets of energy $E=h\nu$.
The delocalized nature of photons in free flight and their localized character in the detection process is at the heart of their wave-particle duality and  very incompatible with classical reasoning.
Here, $\nu$ relates the frequency of the
electromagnetic wave to the energy $E$ of each individual photon
through Planck's constant of action $h$.

The superposition principle is also associated with the dualism between determinism and quantum randomness: The shape and
location of the envelope of the interference pattern is strictly
determined by a wave equation. The square
of the wave amplitude represents the probability to
detect a single event. Each individual realization is, however, objectively random within the predetermined probabilities.

Quantum theory describes the evolution of delocalized objects
perfectly well, but a measurement signifies a break in
this evolution and introduces a random element. Since it
terminates the wave-like propagation of the quantum state it is
often described as a collapse of the wave function. It is still
an open debate whether measurements have an ontological or
rather an epistemological meaning, i.e. whether they describe
an outside reality or our knowledge about the world. But for many practical purposes the act of measurement is a valid and useful concept.

It is important to note that interference always occurs, when the wave associated to a quantum object may reach the detector along at least two different but intrinsically indistinguishable paths - either in real-space or in some configuration space (e.g. potential curves).
The Mach-Zehnder interferometer, sketched in Fig.~\ref{Quantum}d, is a simple textbook arrangement, which is often used to demonstrate delocalization and interference for photons over macroscopic distances. It also visualizes an idea that has been invoked in the description of coherent energy transport in organic molecules.

\subsubsection{Quantum delocalization of matter}
Diffraction and interference deviate even more from our classical expectations when we observe them with massive particles, such as electrons, neutrons or atoms~(\citealp*{Cronin2009a}, and citations therein). But chemistry teaches us that the delocalization of electrons over nanometers is rather ubiquitous, for instance across the bonds of aromatic molecules.

The wave-nature of matter is also of relevance for life science in
a different context: The short de Broglie wave length of fast electrons allows to obtain high resolution images in electron microscopy and
neutron waves are useful in analyzing crystallized protein structures.

While electrons and neutrons are still rather small and elementary particles,
recent experiments were able to demonstrate the delocalization
of entire molecules, such as C$_{60}$, using far-field diffraction~\citealp*{Arndt1999a}.

\begin{figure} [t]
 \begin{center}
 \includegraphics[width=0.7\columnwidth]{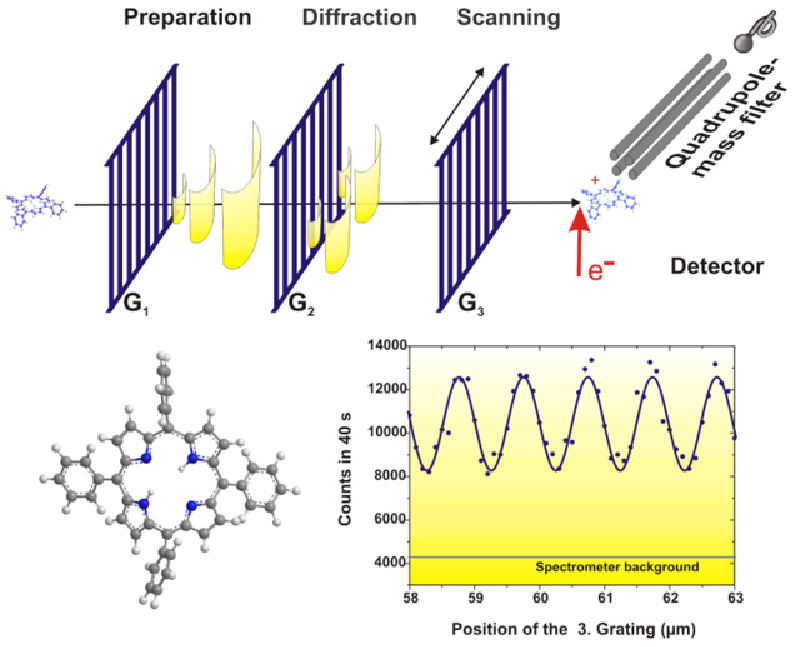}
  \caption{The wave-particle duality of the biodye
 tetraphenylporphyrin can be revealed by diffracting the
 molecules in a near-field interferometer of the Talbot-Lau
 type~(\citealp*{Hackermuller2003a}). Molecules passing
the first grating are diffracted and
 delocalized over several micrometers at the second grating. Diffraction there leads to interference fringes, i.e. a molecular density pattern, at the position of the third mask. This is imaged by
 scanning grating G$_{3}$ and by recording all transmitted molecules in a mass spectrometer.}\label{TPPinterference}
 \end{center}
 \end{figure}

And even biodyes, such as  porphyrin derivatives, revealed their quantum wave properties in a near-field interferometer~(\citealp*{Hackermuller2003a}), as shown in
Fig.~\ref{TPPinterference}) and briefly outlined below: the molecular powder was evaporated to form a beam, which was velocity selected to
$\Delta v/v\simeq$ 15--20\,$\%$. The stream was directed onto a gold grating, i.e. an array of slits (450\,nm wide) with a period of 990\,nm. The molecules are only nanometers large and their center-of-mass is still localized to within a few picometers inside the oven. But once ejected into the vacuum, the free quantum propagation stretches their coherence function by more than a factor of 10,000, i.e. to 50\,nm before they interact with the first
grating. Diffraction within the slits of this grating
leads to a further broadening of the molecular coherence function, up to several micrometers before the porphyrins encounters the second grating. The simultaneous and indistinguishable passage of the molecular wave amplitude through several slits leads to quantum interference, which manifests itself as a sinusoidal molecular density pattern at the location of the third grating (Fig.~\ref{TPPinterference}).

These experiments show that large scale coherence of biomolecules is obviously observable. These molecules can be delocalized over many
micrometers, thousand times their physical size and million times their de Broglie wavelength, and over time scales of a few milliseconds. But it is also clear that this demonstration required laboratory conditions, i.e. high-vacuum to avoid observing interactions with the environment.
It is interesting to note that all porphyrin molecules were heated to 690\,Kelvin and thermal excitations could drive internal state changes even during the molecular propagation through the interferometer. The reason for de Broglie interference to survive lies in the fact that at this temperature the internal states still remain effectively decoupled from the center-of-mass motion~(\citealp*{Hackermuller2003a}).

Several groups have formulated serious research proposals that ask to which extent quantum superpositions might still be observable on the mass scale, not only of a single organic molecule, but possibly even of a supermassive cluster or a virus (\citealp*{Clauser1997a,Reiger2006a,Romero-Isart2009a}). Chances are high that the idea behind these proposals can actually be successfully implemented within the next decade.

\subsubsection{Quantum tunneling}
When a particle encounters a potential barrier that is higher
than its kinetic energy, it will not be able to pass by any
classical means. Quantum mechanics, however, allows it to tunnel
through the barrier (see Fig.~\ref{Quantum}c).
In order to understand the relevance of this phenomenon for biology it is important to see that the tunneling probability depends exponentially
on the height, width and shape of the potential barrier.
The particle's mass and kinetic energy enter in a similar way. This explains
why tunneling of electrons and protons is
rather ubiquitous in biology whereas it is unobserved for entire amino acids or proteins:  their mass and their interaction potentials grow with the number of molecular constituents and therefore dramatically decrease the probability for this quantum effect.

\subsubsection{Spin: a quantum way to turn around}
Angular momentum $L$ is a well-established quantity in classical
physics, but quantum physics requires that
it appears only with a discrete set of values and orientations.
Most particles also possess a spin, i.e. an angular momentum that is not related to any mechanical motion and therefore without any classical analog.

Spins may also be brought into superposition of two or more  mutually exclusive states, comparable to a single compass needle pointing both north and south at the same time. In spite of being a pure quantum property, spin is also often responsible for magnetism in biological systems on the molecular level. The spin of protons is also exploited to derive structural and functional properties of organisms~(\citealp*{Lauterbur1973}).
Magnetic resonance imaging of human tissue relies on the  simultaneous response of many trillion spins in a macroscopic biological volume.
It is, however, still an open question whether magnetization in biology can be quantized on a larger scale, whether spin entanglement or mesoscopic coherent spin transport can be found under ambient conditions and with a consequence for biological functionality - a question that is resumed below in the context of the magnetic senses in animals.

\subsubsection{Quantum superposition of energy states}
If a molecule is excited by a femtosecond optical pulse, the energy-time uncertainty relation
$\Delta E \times \Delta t \ge \hbar/2$ ensures that the photon
energy is sufficiently undetermined to be resonant with more than
a single excited state. Femtosecond spectroscopy has thus become a regular tool for characterizing biomolecular
systems~(\citealp*{Felker1982}) and the coherent superposition of electronic and vibrational energy states are regularly observed in such experiments.
But even when biological systems are exposed to incoherent daylight, excitonic coherences may form between electronically coupled neighboring molecules. In photosynthetic complexes excitonic coupling across several pigment molecules has been reported~(\citealp*{Dahlbom2001a}) and coherence in photosynthesis is a major field of current research (see further below).

\subsection{Quantum statistics}
The spin of identical particles also determines their statistical
behavior within an ensemble. Bosons, particles with an integer value of the spin quantum number $s$ tend to occupy the same quantum state when they are prepared to be indistinguishable.  Fermions with  half-integer quantum numbers, such as electrons, avoid each other under otherwise identical circumstances. This has
important consequences: Pauli's exclusion principle, the stability of matter and the existence of neutron stars are important results for fermions. On the other hand, the existence of lasers and Bose-Einstein condensates are macroscopic effects of boson statistics. The observation of quantum  statistics with atoms requires highly specialized environments, such as $\mu$K temperatures and ultra-high vacuum. Interestingly, quantum degeneracy of exciton-polaritons has already been observed in a condensed matter system at T=19\,K~(\citealp*{Kasprzak2006a}). This is much closer to the conditions of life, but at present there is still no experimental evidence for  quantum statistics on a macroscopic biological scale.

On the other hand, it is worth noting that photons are also bosons with a spin quantum number of $s=1$, and non-classical states of light may well be generated in organic systems. Although we don't know of any biological
laser, random lasing of artificially dye-infiltrated human
tissue has been demonstrated, recently~(\citealp*{Polson2004a}).

\subsection{Quantum entanglement and quantum information}
When we extend the superposition principle to more than a single
object or property, we arrive at the notion of quantum
entanglement. By this we mean a non-classical correlation, an
unseparable connection between two objects or properties.
Amazingly, once established, this quantum connection may
theoretically persist over long distances and times, unless it is
perturbed by external interactions and measurements.

The example of polarization entangled photon-pairs shall
illustrate the idea (Fig.~\ref{Quantum}f): When photons are
sent onto a properly chosen non-linear crystal,
 each of them can be converted with a certain probability into two photons of lower energy. The polarizations of the two photons is then quantum correlated, i.e. entangled, in the following sense:
 none of the two emerging photons has a well-defined
polarization before it is measured, but when the two quanta of light are polarization-entangled we can predict in advance and with certainty that the polarization of both will always be orthogonal to each other. If the first photon is
horizontally polarized, denoted as $|H\rangle$, the partner quantum will be detected in a vertical orientation, denoted as $|V\rangle$, and vice versa.
This is why quantum physicists write this particular entangled state as $\psi \propto |H_{1} \rangle | V_{2}\rangle \pm | V_{1}\rangle | H_{2}\rangle$.  The sum or difference of the two products describes a quantum superposition of the two possibilities. The essence of quantum entanglement lies in the  fact that nature no longer allows us to describe one particle without the other. Their individual polarization is not defined until we do a definite measurement. Each individual recording will find a random orientation for each individual photon, but if we compare the properties of paired quanta, they will always be perfectly correlated.

There are several formal definitions and measures of
entanglement~(\citealp*{Vedral1998}). Generally the following
conditions are required to test for it:
First, we have to identify at least two physical subsystems or
modes. They must be clearly distinguishable and independently
addressable for instance through different frequencies, locations,
or detector orientations. Secondly, we have to be able to identify at least two complementary measurements that can be performed on each of these
two modes, and it should be possible to gradually vary between the
two kinds of measurements.
In the photon example, these two criteria are easily fulfilled as
the two polarizations are orthogonal and optical
elements such as $\lambda/2$-plates can transform one into the
other. The two twin-photons can be emitted in spatially
separated arms, too.

Quantum entanglement has been demonstrated for photons, but also for superconducting circuits, nuclear spins in small molecules, spin noise in atomic ensembles, trapped ions and other systems~(see refs. in \citealp*{Nielsen2000a}).
All these proof-of-principle demonstration experiments were, again, performed under strictly controlled
environmental conditions, often including ultra-high vacuum and
ultra-low temperatures -- i.e. conditions
that are incompatible with living organisms.

Interestingly, entanglement may also be mediated by the exchange
of light. And the transmission can be amazingly robust, as shown
by the successful quantum teleportation
of atomic spin properties from one room temperature atomic ensemble to
another, over macroscopic distances through
air~(\citealp*{Sherson2006a}). Electromagnetic radiation might thus be an important coherence mediator, if entanglement should be relevant in life.

Quantum correlations are of particular interest for new information processing schemes~(see refs. in \citealp*{Bouwmeester2000a,Nielsen2000a}): in contrast to classical physics, where a computing bit can only assume the
values  0 or 1, quantum systems can coexist in a superposition of states and form quantum-bits or qubits of the form $\psi \propto |\,0 \rangle +
e^{i\phi} |1\rangle $, where $\phi$ is an angle that can take a continuous, i.e. infinite, range of values. This continuity opens, in principle, a way to massively parallel computation.

As of today, several algorithms have been suggested which would actually provide a significant advantage over all classical schemes, when
implemented with many qubits~(see refs. in~\citealp*{Nielsen2000a}). Quantum
computers are expected to provide faster prime number factoring
for crypto-analysis using Shor's algorithm, speed-up in database search using Grover's algorithm, new insights into quantum games~(\citealp*{Eisert1999a}) or an exponential speed-up in solving systems of linear equations~(\citealp*{Harrow2009a}).

These developments in quantum information science lead to the question whether quantum methods may also be relevant for living organisms, which are stunning information processing devices.
Yet, the implementation of quantum circuits in biology, as anywhere else, would require a configuration space that increases exponentially
with the number of qubits involved. And given the enormous
sensitivity of quantum states to most external perturbations it is widely believed that the probability of a successful coherent computation
will also be exponentially suppressed under the conditions of life as they are known to us.

Until today, first quantum computing circuits and elementary quantum algorithms could be demonstrated using nuclear magnetic resonance (NMR) on a few nuclear spins {\em within} an individual biomolecule~(\citealp*{Jones2000a}, see Fig. 3a).
These experiments approach a natural setting since they were embedded in a condensed matter environment, even though
at a temperature of a few Kelvin, only. NMR quantum computing in a dense ensemble is mostly  hampered by the difficulty to initialize all
molecules to the same ground state. And also the rapid randomization of quantum phases by interactions with the host matrix is a further limiting factor.

One may also speculate whether naturally occurring quantum correlations
can be used in biology.  Being embedded in a macroscopic system, any initial coherent superposition of a molecule -- be it in position, electronic
or vibrational excitations --  will rapidly be transformed into entanglement with the environment. One could imagine that this yields a functional
benefit. But again, experimental evidence for this and other meaningful quantum information processing in nature still has to be found.

\section{The quantum-to-classical transition}

\subsection{Some general insights}
In order to understand the overwhelming success of our classical world-view in the description of biological phenomena we have to understand, why quantum effects are usually hard to observe.

The {\em kinematic argument} is based on the insight that Planck's
constant of action $\hbar$ is extremely small. For instance, the
de Broglie wavelength $\lambda_{dB}=h/mv \simeq 10^{-35}$\,m  of
an adult man walking at a speed of 1\,m/s is way too small to
ever be observed. Here, it is rather the value of the fundamental constants than the intrinsic structure of quantum mechanics which forbid the observation of macroscopic quantum delocalization.

The {\em phase averaging argument} adds the insight that
quantum interference relies on phase coherence. But small
wavelengths are easily dephased in changing environments.
Fluctuations of the geometry, electromagnetic fields or chemical
environments will easily alter the conditions for constructive and
destructive interference. In most cases thermal fluctuations will render
interference phenomena unobservable, when we look at macroscopic
molecular ensembles.  In fact, this does not necessarily mean that quantum
mechanics does not play a role in biology: the stochastic dephasing of destructive quantum interference has recently been even invoked
as being responsible for the fast energy transport in the photosynthetic complex~(\citealp*{Caruso2009}), as further described below.

In contrast to the notion of phase averaging, we here reserve the term {\em decoherence} for processes which actually exchange information between the quantum system and its
environment~(\citealp*{Zurek1991a,Joos2003a}). Decoherence is closely
related to the {\em act of measurement}. By definition, a quantum
measurement extracts information about a quantum
system and correlates it to the pointer of a
macroscopic meter. But in contrast to the prescriptions of unitary quantum physics, we never see any macroscopic meter pointing up and down at the same time.

A valid explanation for that is given by decoherence
theory~(\citealp*{Joos1985a}): The act of measurement on a
quantum superposition state will create entanglement between the
system and the meter. The quantumness of the original system is
thus diluted to a larger composite system (object and meter)
and in most cases, the quantumness of the meter diffuses over
the entire detector which may contain some $10^{20}...10^{25}$
atoms. This is such an incredible amount of particles, that
we will never be able to retrieve the original quantum coherence from
the enlarged system.

In contrast to decoherence theory, for which the unitary
evolution of quantum physics always persists -- even during a
measurement -- the model of an {\em objective collapse of the
quantum wave function} assumes that coherences spontaneously
disappear in the act of measurement. Spontaneous collapse models
assume that this intrinsic loss of coherence is bound to a
certain internal complexity and number of participating particles
(\citealp*{Ghirardi1986a}). Recently, it has also been suggested that
the inevitable coupling between mass and deformations of spacetime
might cause an objective collapse of the
wavefunction~(\citealp*{Diosi1989a,Penrose1996a}).

The kinematic argument, phase averaging and decoherence are all
consistent with the unitary evolution of quantum physics. They
are experimentally accessible and proven to be relevant in
well-controlled experiments. But none of them includes a
qualitative transition between quantum and classical phenomena -
only a gradual reduction of the observability of quantum effects.
In contrast to that, collapse models, as taught in many modern
physics textbooks, postulate a factual and abrupt change between
quantum and classical dynamics.  But also this assumption leaves
many experimental and philosophical questions unanswered.
Also from this fundamental point of view, it is therefore still open to which degree quantum physics can prevail on the macroscopic scale of living species.

\subsection{Can quantumness survive in biological environments?}

The observation of quantum phenomena often requires low
temperatures and a high degree of isolation.
And yet, quantum coherence could be observed in matter-wave experiments with
biodyes even at {\em
internal} temperatures exceeding 690\,K~(\citealp*{Hackermuller2003a}).
Related experiments, however, also quantified the high sensitivity of quantum coherence to the emission or scattering
of even single photons~(\citealp*{Hackermuller2004a})
 or molecules~(\citealp*{Hornberger2003a}).
A simple and general rule reads as follows: {\em If an individual state of a quantum superposition can be resolved and detected by any interaction with its environment, the superposition will be destroyed.} A practical example: A molecule that coherently traverses several slits of a diffraction grating will maintain its quantum superposition until the environment gets the capability of reading which slit the particle takes. Even if the openings are separated by only  $250$\,nm, the scattering of a single visible photon per molecule may destroy the observation of quantum interference.

Under biological conditions, in an aqueous and warm environment, position superpositions of massive
particles cannot survive more than a few collisions with electrons, atoms, photons or phonons. In the case of strong interactions, a single collision will destroy the coherence. And even for weak perturbations, the visibility of quantum phenomena will vanish exponentially with the separation of the position states. Any delocalization in biomaterials will therefore be limited to extensions of a few nanometers and time scales typically shorter than nanoseconds, in most cases even a few hundred femtoseconds, only.  But a few backdoors still exists for coherence and entanglement to persist:

First, it is conceivable that special molecular architectures may shield some parts of a system from some interactions with
their environment. Electrostatic forces may for instance create hydrophobic pockets from which the solvent could be excluded. But  the existence and survival of {\em decoherence free subspaces}~(\citealp*{Lidar1998}) in organic matter and over the time scales of milliseconds still has to be shown  experimentally.

Secondly, {\em quantum error correction}~(\citealp*{Shor1995a}) is
often being discussed as a loophole for the implementation of quantum information processing in living systems.
The idea is based on the fact that a certain class of computational errors may even be corrected if no one knows the details of the state or the errors that occurred. This is particularly important, as any attempt to read a quantum superposition would result in its loss, i.e. its reduction to a classical state.
A system with quantum error correction requires, however, even more qubits than the uncoded counterpart. The idea of the scheme has been demonstrated with a 3-bit code in NMR quantum computing with alanine~(\citealp*{Cory1998a}), but as of today, extensions to more than a handful of spins at low temperatures are not known.
We thus face the following dilemma: What are the odds that quantum error correction was developed by natural selection from unprotected quantum computing if this faulty precursor process would provide no evolutionary advantage?

Finally, it is worth noting that biology operates in open systems,
i.e. far from thermal equilibrium. It has been
hypothesized that living systems might act as heat engines, providing
{\em local cooling} in an otherwise thermal environment
~(\citealp*{Matsuno1999}).
Another perspective was opened~(\citealp*{Cai2008}) with the hypothesis that entanglement might be regularly refreshed in a thermally driven, repetitive contact between neighboring molecular subsystems. If that happens on a time scale shorter than the one of decoherence, net entanglement may possibly persist.
Clearly, all of these ideas are presently of an exploratory nature.

\subsection{How can quantum entanglement be revealed in mesoscopic systems?}
We introduced entanglement as the inseparability of quantum states. But how could we possibly detect this intricate quantumness in a complex condensed matter environment, where there is no hope of getting access to all underlying details?
Interestingly, it has been found, that even macroscopic thermodynamical variables may serve as entanglement witnesses~(\citealp*{Vedral2003a,Amico2008}).
And to a certain degree, entanglement may even survive in a macroscopic thermal state~(\citealp*{Markham2008a}).

Since the Hamiltonian, i.e. the operator of the system's total energy, that is used for deriving all thermodynamic functions is influenced by the entanglement of the underlying wavefunctions, the energy terms and therefore also all  thermodynamic variables are affected by these quantum correlations.
A single macroscopic observable can thus suffice to hint to the presence of quantum entanglement and the important trick is to find an
observable that can be reliably read in experiments.
In practice, the magnetic susceptibility~(\citealp*{Brukner2006a}) or the heat capacity might provide good indicators for entanglement~(\citealp*{Wiesniak2005a}) and it will be interesting to search for further witnesses in biological systems, too.

There is an enormous reduction of information when we derive a single thermodynamical variable from thousands of quantum states. Entanglement witnesses therefore cannot tell us all details about a system. They rather indicate under which conditions a state is definitely entangled, but they neither quantify it nor do they exclude with certainty the presence of entanglement if the witness conditions are violated.

\section{Life science research with an interface to quantum physics}

\subsection{Single-photon phenomena in the life sciences}
Single quanta of light have been relevant for illustrating
fundamental quantum principles but they are also ubiquitous in
the life sciences: The most efficient detection techniques
for fluorescent biomolecules are sensitive on the single photon
level.
Individual particles of light are also of direct relevance in biological processes as they may affect the structure of individual molecules
which in turn can transduce signals in living
organisms. The retinal molecule can switch its conformation after absorption of very few photons and thus turns the human eye into one of the most
sensitive light detecting devices that exist. Between two and
seven photons are usually sufficient to be perceived by a
dark-adapted human observer~(\citealp*{Hecht1942}).
Various studies indicated that test persons could even count the number of photons with a reliability that was only
limited by quantum shot noise~(\citealp*{Rieke1998}, and citations therein).

Single photon detectors are of great interest for
quantum communication and it has recently
been suggested that octopus rhodopsin, chosen by evolution because it is well-adapted to the dark of the deep oceans, may be a useful component in
such applications~(\citealp*{Sivozhelezov2007}).
But also single photon sources are gaining increasing importance in quantum communication or computation protocols and single molecules are considered to be relevant emitters~(\citealp*{Lounis2005a}).

When we talk about the quantum properties of light we usually
refer to its wave-particle duality, the graininess and
quantum statistical properties, such as photon bunching,
anti-bunching or squeezing~(\citealp*{Glauber2006}). Fluorescence
correlation experiments with proteins~(\citealp*{Sanchez-Mosteiro2004})
have revealed both the discrete quantum nature of
molecular energy states and the non-thermal quantum
statistics of light: An excited single molecule
may usually not absorb a second photon of identical wavelength - unless the excited state has decayed. The emitted photons are therefore released with a time structure which differs from that of thermal light sources.
Photons emitted by a single molecule come in an anti-bunched rather than a bunched time series.  It remains, however, still open whether single-photon emission is explicitly used by living systems.

In contrast to that, artificially grown quantum emitters have
found many applications in the life sciences. The
characteristic energy of a quantum system is connected with its
spatial dimensions. This is in particular also true for semiconductor
nanocrystals, which measure only a few nanometers in diameter and
whose color can be changed from blue to red by growing them to
larger sizes. Fluorescent quantum dots are used as highly
efficient labels for biomolecular imaging and they allow to follow the dynamics of marked receptors in the neural
membrane of living cells~(\citealp*{Dahan2003a}). Similar results have recently also been achieved with nanodiamonds.
Their nitrogen vacancy centers exhibit strong and stable fluorescence, they are biocompatible and they have also been proven to be highly sensitive
quantum probes for magnetic fields on the nanoscale ~(e.g. \citealp*{Balasubramanian2008a}).

\begin{figure}[tbh]
 \begin{center}
  \includegraphics[width=0.9\columnwidth]{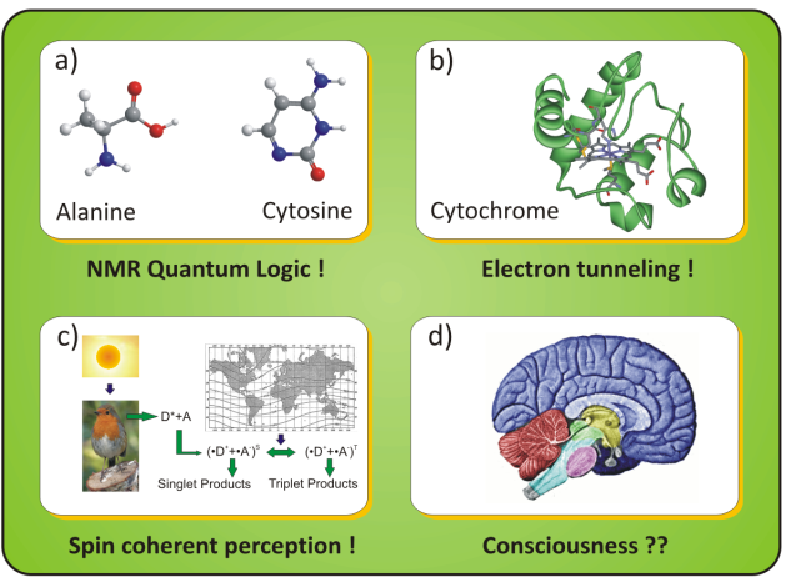}
  \caption{Some recent explorations of quantum aspects in the life sciences: (a) The nuclear spins of amino acids have been used as qubits in quantum computing demonstrations~(\citealp*{Jones2000a}). (b) Electron tunneling on nanometer scales has been established as a common phenomenon in life, for instance in reactions with cytochrome~(\citealp*{Devault1966a}). (c) Electron spin entanglement and coherent spin transport is part of a possible explanation for the magnetic orientation of migratory birds ~(\citealp*{Ritz2000})(d) Speculations about the influence of quantum physics on human consciousness are often regarded as inspiring but as of today they are not substantiated by any experiment (Robin picture: David Jordan, CC-BY-SA). }\label{Bioquantum}
 \end{center}
\end{figure}

\subsection{Quantum tunneling in biomolecules: from enzymatic reactions to the olfactory sense?}
Living organisms are enormous biochemical reactors, making and
breaking zillions of chemical bonds every day. To a large extent
the reaction rates are controlled both by thermal activation and
enzymatic catalysis.  It has been a long-standing question
whether quantum tunneling is also involved and whether its presence
provides  an evolutionary advantage.
This concerns the tunneling of electrons, protons and even entire small molecules.

The theory of electron transfer has a long
history~(\citealp*{marcus1956}). First evidence
for {\em electron tunneling} was derived from the oxidation rate
of cytochrome (see Fig. 3b)in the bacterium chromatium vinosum. Since the
reaction speeds were both large and temperature independent at
low temperatures ($<$100\,K) it was concluded that they are incompatible
with a thermal activation model alone~(\citealp*{Devault1966a}).
Electron tunneling has actually been identified as a widespread process, found in photosynthesis~(\citealp*{Blankenship1989}), cellular respiration~(\citealp*{Gray2003}) and electron transport along DNA~(\citealp*{Winkler2005a}).

While speculations about {\em proton tunneling} had also been
around for long~(\citealp*{loewdin1963}), first experimental evidence
was only given in 1989~(\citealp*{Cha1989a}) for  the enzyme alcohol
dehydrogenase, which transfers a proton from alcohol to
nicotinamide adenine dinucleotide. Since tunneling depends
on the mass of the object, the tunneling rates must change when
hydrogen is replaced by the chemically equivalent deuterium which
doubles the atomic mass. This kinetic isotope effect was
confirmed and gives good evidence for the presence of proton
tunneling. Since then, many other enzymatic reactions were
ascribed to proton tunneling ~(\citealp*{Glickman1994}).
It has to be noted, however, that the tunneling distances involved in all these reactions are typically shorter than 0.1\,nm and the protons traverse the barrier at energies around 10\,kcal/mol (0.4\,eV) below the potential maximum~(\citealp*{Masgrau2006}).

The {\em simultaneous tunneling of several particles} has also
been discussed, including double, triple and even quadruple proton
exchange in cyclic molecular networks~(\citealp*{Brougham1999a}).
The transition rates in these experiments were measured using NMR spectroscopy and the temperature dependence of the reaction rate as well as the kinetic
isotope effect were taken as witnesses for the presence of
hydrogen tunneling.
Even the {\em tunneling of entire small molecules}, i.e. formaldehyde (CH$_2$O), was proposed based on the temperature dependence of its photo-induced polymerization rate~(\citealp*{Goldanskii1979a}).

Turin~(\citealp*{Turin1996}) also opened a public debate by
suggesting that we are even able to {\em smell quantum tunneling}.
Most aspects of our sense of smell are very well understood without it. Linda Buck and Richard Axel received the Nobel prize for their
description of the mammalian olfactory system. They identified transmembrane proteins that encode for odor receptors in the olfactory epithelium ~(\citealp*{Buck1991}). Each of them can sense multiple odorants. And each
odorant can be detected by different sensors.
Most smells can be perfectly explained~(\citealp*{Zarzo2007}) by assuming a lock-and-key mechanism, where an odor molecule binds to a specific receptor combination depending on its size, shape and chemical groups.

Based on much earlier hypotheses~(\citealp*{Dyson1938}),  Turin
suggested that smell is, at least additionally, correlated to the
vibrational spectrum of molecules and that the receptors
perform phonon-assisted inelastic electron tunneling
spectroscopy to identify the odorant.
This idea should explain why our nose is able to distinguish molecular groups of similar geometry but different vibrational spectra, such as for example OH and SH or ferrocene and nickelocene~(\citealp*{Turin2002}).
However, recent experiments~(\citealp*{Keller2004}) rejected this theory, while newer theoretical work conceded a conceptual viability of the idea -- even though without being quantitatively decisive~(\citealp*{Brookes2007}).

Concluding, we see that quantum tunneling is certainly present in a
large number of biological processes, but experimentally proven only on
the level of small-scale chemical reactions.

\subsection{Coherent excitation transfer in photosynthesis}
Photosynthesis is a key process for life and often considered as a role model for future light harvesting technologies~(\citealp*{Blankenship2002}).
It is differently realized in plants, algae or bacteria. But they all convert light to chemical energy. A closer look reveals that photosynthesis involves a plethora of highly complex processes, such as long-ranged excitation transfer, redox-reactions, hydrolysis, proton transport or phosphorylation.
In many parts of the system -- including the wet-chemical material transport -- we don't expect to find significant quantum coherence or entanglement, but
others may actually require the notion of quantum tunneling, coherent excitonic transfer or matter-wave interference.

The photosynthetic complex is a membrane-bound system with many
embedded functional subunits. The energy conversion starts with the absorption of an incident photon by a pigment molecule, e.g. a chlorophyll, porphyrin  or a carotenoid molecule embedded in a protein structure, the antenna complex.
The large number of these dye units ensures a high
photoabsorption probability, and their arrangement
enables an efficient excitation transfer from the primary absorber to the reaction center.

The reaction center is a pigment-protein complex which contains a dimer, called the special pair. When it is excited, it donates an electron to a neighboring acceptor molecule. Fast secondary processes prevent the recombination of the ion pair and trigger the release of protons that are first transferred across the membrane and later used to fuel, for instance, the synthesis of adenosine triphosphate (ATP) from adenosine diphosphate (ADP).

\begin{figure} [tbh]
 \begin{center}  \includegraphics[width=0.6\columnwidth]{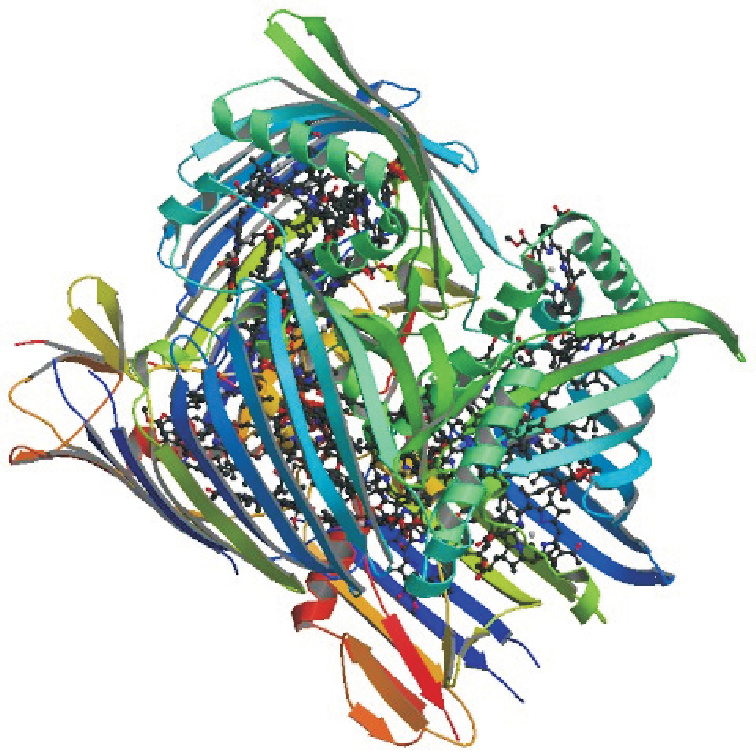}
  \caption{The Fenna-Matthews-Olson (FMO) complex is composed of three protein-pigment structures. Each of them contains seven bacteriochlorophyll-a molecules~(\citealp*{Blankenship2002}).  Electronic excitation transfer from the FMO complex to the reaction center is a key process in the light-harvesting of green photosynthetic bacteria. Two dimensional Fourier-Transform spectroscopy~(\citealp*{Engel2007}) was able to document long-lived excitonic coherences across neighboring molecules in this structure (picture credits:~\citealp*{Tronrud2009a}).
  }\label{Photosynthesis}
 \end{center}
\end{figure}

 Several recent studies~(\citealp*{Grondelle2006,Cheng2009})
emphasized how well the excitation transfer from the antenna pigments to the reaction center is optimized. A fast conversion is important since any delay would increase the chances of relaxation mechanisms to channel energy into heating instead of chemical potentials.

Early explanations of the energy transport, based on incoherent and dipole-dipole-mediated excitation hopping between molecular sites~(\citealp*{Foerster1948}), failed to explain the observed transfer rates. Delocalization and coherent exciton coupling between the closely packed antenna pigments were therefore suggested as the most likely explanation, with experimental support rapidly growing throughout recent years.

Modern two-dimensional Fourier transform spectroscopy allowed to probe the various excitation transfer pathways between the molecules on a femtosecond time scale.
In particular, experiments performed on a 77\,K cold bacteriochlorophyll
Fenna--Matthews--Olsen (FMO, Fig.~\ref{Photosynthesis}) antenna complex
 were able to reveal exciton delocalization~(\citealp*{Brixner2005}) and long-lasting coherence in the excitonic energy
transfer~(\citealp*{Engel2007}).

The observation of a spatially and temporally extended coherence, covering several nanometers and time-spans as long as a few hundred femtoseconds, is highly remarkable and it has triggered a growing number of scientific groups to focus their theoretical and experimental work on that question.
As of today, a rich set of detailed data has already been collected to characterize the energy levels, transfer rates, intramolecular and intermolecular coherences. In particular the latter raised the question
how to connect these findings to related fields in quantum physics.

When there is coherence,  what is the role of constructive or destructive interference? And are we allowed to use the language of quantum information processing to describe the highly efficient natural transfer of information and energy in light-harvesting complexes?    
It has been suggested that a 'wavelike' sampling of
the energy landscape or even a quantum search algorithm might permit to
find the fastest route from the antenna to the reaction
center~(\citealp*{Engel2007}).
The excitation transport has also been associated with quantum random walks~(\citealp*{Mohseni2008a}). In contrast to classical random walks - which we also know from Brownian motion - the position of the quantum walker would not be a single random position but rather a superposition of positions.

The incorporation of interference effects in the theoretical reasoning led to further considerations concerning the possible role of the protein environment ~(\citealp*{Rebentrost2009,OlayaCastro2008}), since a close look at wave physics reveals that coherence can be both beneficial and a hindrance if the aim is to optimize the speed of transport. On the one hand,  the simultaneous wavelike sampling of many parallel paths could possibly result in finding a faster way to the final goal. But on the other hand
the presence of an irregular lattice of scattering centers (static disorder) may actually suppress wave transport because of destructive interference. This phenomenon, well know in solid state physics, is called Anderson localization~(\citealp*{Anderson1958a}). In that case, thermal fluctuations of the protein environment might therefore be crucial and help to avoid localization and thus assist in the excitation transfer~(\citealp*{Caruso2009}).
The importance of protein dynamics in eliminating Anderson localization was actually already discussed in an earlier paper by~(\citealp*{Balabin2000}), where multiple quantum pathways and interference were proposed for the {\em electron transfer} after the reduction of the special pair -- instead of the excitation transfer towards the special pair that is discussed here.

The role of interference in transport phenomena can also be visualized by recalling the analogy to an optical Mach-Zehnder
interferometer (as shown in Fig.~\ref{Quantum}d): Depending on the setting of phases, wave interference can guide all excitations to either one of the two exits.  Quantum coherence may then be the best way to
channel the interfering quanta to the desired output. But if the wave phases
happened to be initially set to destructive interference, quantum coherence
would be a severe handicap. In this case, even random dephasing processes would help to optimize the transport efficiency.

External perturbations may also be important for energetic reasons: the electronic excitations have to be transferred between complexes of
different energy. If the molecular states were too well defined,
the lacking energy overlap would reduce the transfer rate. External perturbations may broaden the transition bands and thus increase the coupling between neighboring molecules.

Recent experiments by (\citealp*{Collini2009a}) however hint also at another possible role of the protein environment. In their experiments they could show that coherent electronic excitation transfer along conjugated polymer chains occurs even at room temperature. These long lasting coherences (200\,fs) could only be observed in intrachain but not in interchain electronic excitation transfers.

All of the models described above bear in common that they rely on quantum
coherence and decoherence and that they may be robust even under
ambient environmental conditions -- over short time scales.
It is thus the fine interplay of coherent exciton transfer, decoherence and dephasing that yields the best results and which seems to reign one of the most important reactions in nature.

\subsection{Conformational quantum superpositions in biomolecules}
Since atoms can exist in a superposition of position states this
may also lead to a superposition of conformational states in molecules.
A tunneling-induced superposition of conformation states is
conceivable. It becomes,
however, highly improbable when many atoms have to be shifted over
large distances and across high potential wells during the state change.

Photoisomerization is another way of inducing structural state
changes in molecules - now using photon exchange, instead of
tunneling. This opens the possibility to connect even energetically separated states. The photo-induced all-trans--13-cis transition of retinal is
a famous example where a single photon can cause a sizeable conformation
change. But much of the subsequent atom rearrangement
occurs in interactions with the thermal
environment~(\citealp*{Gai1998}).
In spite of that, it was possible to gain coherent quantum control in this process. Applying
pulse-shaped femtosecond laser excitation to retinal in a native protein
environment~(\citealp*{Prokhorenko2006}) achieved a modulation of the
isomerization yield by $\pm 20\,\%$. The detected dependence
on the laser phase is a good indication for the relevance
of quantum interference among vibrational states.
But a coherent superposition of functionally different configuration states, instead of electronic or vibrational states, has not been achieved for any large biomolecule, so far.

Decoherence has often been named to explain the prevalence of chirality in biomolecules. If a molecule may exist in two enantiomers, quantum mechanics allows, in principle, also for a coherent superposition of the left-handed and the right-handed state. In practice, however, this is not observed for larger particles.
An intuitive argument is based on the fact that various scattering processes between a molecule and its environment depend on its chirality. This may include the scattering of polar light and elementary particles or the interaction through higher-order London dispersive forces between polarizable bodies. Such events may act as quantum measurements and projections onto a chirality state. And in many cases, the energy barrier between the symmetric ground states will then be too high to allow for their spontaneous mixing on a time scale comparable to the scattering events~(\citealp*{Trost2009a}).

The generation and controlled decoherence of chirality superposition states in biological molecules thus still remains an open challenge. The lack of any experimental evidence for coherent conformation superpositions in large molecules also seriously questions a recent model by Hameroff and Penrose who suggested that the collapse of such superpositions in microtubuli may be the cause for the emergence of human consciousness~(\citealp*{Hameroff1996a}).

\subsection{Spin and the magnetic orientation of migratory birds}
It is well established that various animals are able to derive
direction information from the geomagnetic
field~(\citealp{Wiltschko1995,Ritz2000,Johnsen2008}).
Some mammals perceive the Earth's field as a polarity compass, distinguishing north and south, while birds and reptiles rely on an inclination compass that discriminates between polewards and equatorwards and which exploits both the intensity and the gradient of the field.
Interestingly, it could be shown~(\citealp{Wiltschko2006a} and refs. therein),  that the the orientation in the magnetic field requires the presence of visible light beyond a certain photon energy and that an oscillating magnetic field (0.1-10\,MHz) can disturb the bird's senses.

It has therefore been argued that vision-based magnetosensing might be rooted in the light-induced formation of a radical pair~(\citealp*{Schulten1978a}),  a mechanism originally invoked to explain the photochemically induced dynamic polarization in nuclei (\citealp{Closs1969a,Kaptein1969a}):  When light falls onto a donor molecule in the bird's eye, it may excite it to a singlet state (Fig. 3c).
The molecule may then transfer an electron to a neighboring acceptor molecule. The freshly formed pair of radical molecules usually starts in a singlet state (total spin quantum number: s=0), but in the presence of hyperfine couplings with the molecular nuclei it will undergo an interconversion between the singlet and the triplet state (s=1).
Since spin is otherwise rather well protected from environmental influences on a short time scale, it is assumed that the spin pair remains quantum correlated, i.e. entangled in this process. This is also supported by a recent calculation (\citealp*{Rieper2009a}) where even a weak external oscillatory magnetic field noise was admitted and not able to fully destroy entanglement.
The evolution of the electron spins both in the presence of the nuclei and the earth's magnetic field will vary the ratio between singlet and triplet states.  Since many chemical reactions are spin-dependent -- in particular also the back-transfer of the electron from the acceptor to the donor -- the spin evolution should also influence the ratio of molecular products that are finally formed in the bird's eye.
A model for the transduction from the radical pair to the neuronal correlates was proposed by (\citealp*{Weaver2000a}) who also estimated the requirements on the size and the temperature dependence of the system in order to yield a certain sensitivity.

The radical pair mechanism was ascribed to the signalling protein cryptochrome that can be found in the bird's retina (\citealp*{Wiltschko2006a}). Both the electron transfer from a photo-excited flavin adenine dinucleotide along a chain of tryptophan molecules and the reverse recombination reaction are supposed to be sensitive to the geomagnetic field~(\citealp*{Solov'yov2009a}).

The idea is further supported by recent experiments of~(\citealp*{Maeda2008a}) who showed that the radical pair mechanism in the earth's field is actually sufficiently strong to alter the chemical end products in a custom-designed complex that was built from a carotenoid, a porphyrin and a fullerene C$_{60}$.

In order to further corroborate that magneto-sensing is related to quantum-correlated (entangled) electrons, (\citealp*{Cai2009a}) suggested to use a sequence of short radio-frequency pulses to obtain active quantum control over the radical pair spins, immediately after their creation.  Such and related experiments are still required to further elucidate this intriguing phenomenon.

\section{Speculations on quantum information and biology on the large scale}
Most puzzles of quantum physics are related to the way
information is encoded and processed. Some researchers would
therefore demand that quantum biology should be defined by its
use of quantum information.
The present section recapitulates two recent speculations which aim at much larger scales than that of a few molecules.
We clearly state that, as of today, these hypotheses are without
any experimental justification and even disputed on theoretical grounds.
But as some of them have gained rather high popularity in discussions they merit mentioning and brief comments.

\subsection{Quantum physics and the human mind}
About two decades ago, Roger Penrose raised the question whether classical
physics alone could suffice to explain the enormous problem
solving capabilities of the human brain~(\citealp*{Penrose1989}). And he
speculated that a combination of currently
irreconcilable pieces of physics, namely quantum theory and
general relativity, might open a new window to our understanding of human consciousness, i.e. another phenomenon which is hardly understood.

Together with the consciousness scientist Stuart Hameroff he proposed a model,
that assumes that the human mind may exploit at least
two conformations of microtubuli  as values of a quantum bit. The quantumness of the proteins was suggested to solve complex computational problems in the brain while the act of consciousness would be linked to a gravity-induced objective collapse of the quantum wave function~(\citealp*{Hameroff1996a}).

Intriguing as the idea of macroscopic quantum coherence may be, the proposed model hits several hard bounds and controversies:
As of today, no one has ever been able to prepare and characterize a useful coherent macroscopic quantum superposition of two conformations in a macromolecule, not even in the lab.
And even if it existed in nature, decoherence is believed to be orders of magnitude too fast to make it relevant on physiological time
scales~(\citealp*{Tegmark2000,Eisert2007}).

An objective collapse of the wave function is currently also only one of
many models to explain the emergence of classicality from
quantum physics.  The dynamics of the proposed gravitational collapse is neither theoretically understood nor experimentally observed.
It may also surprise that microtubuli were chosen as the decisive agents in quantum consciousness. They are by no means special to the human brain but rather ubiquitous cell support structures.

In spite of its potential deficiencies, the model serves a purpose in that it stretches the scientific fantasy to its very limits. And even though it is unlikely that all details of the proposal will survive future scientific explorations,  experimental efforts in proving or disproving these details will lead to new insights into the relevance of quantum phenomena within the life sciences.

\subsection{May quantum physics speed-up biological evolution?}
The idea starts from the question how a complex protein or strand of DNA could possibly have formed by random trials and mere chance from primordial amino acids or a series of nucleotides up to the high degree of complexity that is required to drive self-replication and evolution.

It has therefore been asked whether a faster, macroscopic quantum sorting mechanism might have been involved in finding the first successfully self-replicating molecule on Earth~(\citealp*{McFadden2000}).
However, its realization on our early Earth must have involved thousands of atoms and molecules in a warm and wet environment under the additional precondition that all sorts of molecules were available, that the formation of the sample molecular structures was energetically accessible and that the molecules were delocalized over large areas in the given environment.

In particular the latter requirement is in variance with the findings of molecule decoherence experiments~(\citealp*{Hornberger2003a}), which confirm that any measurement - be it collision with other molecules, phonons or photons - is capable of destroying the quantum delocalization, if the interaction retrieves position information.
But even if we hypothesize that large molecules could be delocalized in a primordial soup, the fastest speed-up in Grover's quantum search has still only a square-root advantage and the number of combinations is still stupendous.
One  might argue that the initial replicators were extremely tiny and that the first useful molecules for life were only influenced or catalyzed by the replication of the tinier structures. But even then: a feedback loop between biological evolution and the suggested quantum coherence remains highly speculative in the light of present knowledge.

\section{Conclusions}
Quantum physics and the life sciences are both attracting
increasing interest and research at the
interface between both fields has been growing rapidly.
As of today, experimental demonstrations of quantum coherence in biology are
still limited to the level of a few molecules. This
includes for instance all quantum chemistry, tunneling processes, coherent
excitation transport and local spin effects.

In recent years quantum biology has stimulated the scientific reasoning and fantasies and has triggered hypotheses ranging from
exploratory and visionary  over speculative to very likely to be simply wrong.
The current status of research does not always allow to draw a precise
borderline between these classifications. Experimental facts
are largely missing, theoretical understanding is still an
enormous challenge and scientists are arguing both in favor and against various of these ideas.

Fascinating combinations of physics and biology can be understood, already now. We have identified a large number of interconnects  between quantum physics and the life sciences and the status of present experimental skills is great. But the complexity of living systems and high-dimensional Hilbert spaces is even greater.

When we talk about quantum information, the discussion always circles around exponential speed-up. But in living systems any improvement by a few percent might already make the difference in the survival of the fittest.
Therefore, even if coherence or entanglement in living systems were limited to very short time intervals and very small regions in space  -- and all physics experiments up to now confirm this view -- simple quantum phenomena might possibly result in a benefit and give life the edge to survive.

We still have to learn about the relevance and evolutionary advantage of quantum physics in photosynthesis, the sense of smell, or the
magnetic orientation of bird. We still don't know whether quantum
entanglement is useful on the molecular level under ambient
conditions, whether quantum information processing could possibly be implemented in organic systems. We still don't fully understand and appreciate the philosophical
implications of the quantum-to-classical transition even under
laboratory conditions.

We thus conclude that  the investigation of quantum coherence and entanglement in biological systems is timely and important. And it will need even more visions, further refined theories and above all -- a significantly broadened basis in carefully worked out and interdisciplinary experiments.

\section{Acknowledgments}
M.A. would like to thank the FWF for support in the Wittgensteinprogram Z149-N16. V.V. would like to thank EPSRC, QIP IRC, Royal Society and the Wolfson Foundation, National Research Foundation (Singapore) and the Ministry of Education (Singapore) for financial support.

\end{document}